# Electrical polarization switching in bulk single crystal GaFeO$_3$

Maria Biernacka[1], Paweł Butkiewicz[2], Konrad J. Kapcia[3,4], Wojciech Olszewski[1], Dariusz Satuła[1], Marek Szafrański[5], Marcin Wojtyniak[6], Krzysztof Szymański[1]

[1] *Faculty of Physics, University of Bialystok, K. Ciolkowskiego 1L, 15-245 Białystok, Poland*

[2] *Doctoral School of Exact and Natural Sciences, University of Bialystok, K. Ciolkowskiego 1K, 15-245 Białystok, Poland*

[3] *Institute of Spintronics and Quantum Information, Faculty of Physics, Adam Mickiewicz University in Poznań, Uniwersytetu Poznańskiego 2, 61-614 Poznań, Poland*

[4] *Center for Free-Electron Laser Science CFEL, Deutsches Elektronen Synchrotron DESY, Notkestr. 85, 22607 Hamburg, Germany*

[5] *Faculty of Physics, Adam Mickiewicz University in Poznań, Uniwersytetu Poznańskiego 2, 61-614 Poznań, Poland*

[6] *Institute of Physics – Center for Science and Education, Silesian University of Technology, Krasińskiego 8, 40-019 Katowice, Poland*

**Abstract**

The electrical polarization switching on stoichiometric GaFeO$_3$ single crystal was measured, and a new model of atomic displacements responsible for the polarization reverse was proposed. The widely adapted mechanism of polarization switching in GaFeO$_3$ can be applied to stoichiometric, perfectly ordered crystals. However, the grown single crystals, as well as thin films of Ga-Fe-O, show pronounced atomic disorder. By piezoresponse force microscopy, the electrical polarization switching on a crystal surface perpendicular to the electrical polarization direction was demonstrated. Atomic disorder in the crystal was measured by X-ray diffraction and Mössbauer spectroscopy. These measurements were supported by ab initio calculations. By analysis of atomic disorder and electronic structure calculations, the energies of defects of cations in foreign cationic sites were estimated. The energies of the polarization switch were estimated, confirming the proposed mechanism of polarization switching in GaFeO$_3$ single crystals.

## 1. Introduction

The GaFeO$_3$ compound crystallizes in the orthorhombic structure of space group *Pna*2$_1$ (No. 33). The asymmetric unit cell contains two non-equivalent iron (Fe1, Fe2), two gallium (Ga1, Ga2), and six oxygen sites. The symmetry operations acting on the atom at point $(x, y, z)$ transform it into three other positions $(x + 1/2, -y + 1/2, z)$, $(-x, -y, z + 1/2)$ and $(-x + 1/2, y + 1/2, z + 1/2)$ yielding 40 atoms in the unit cell. The atomic positions corresponding to two opposite polarizations of this polar compound are shown schematically in Figs. 1 and 2. The cation in the Ga1 site has tetrahedral coordination, while in the remaining Ga2, Fe1, and Fe2 sites, cations are coordinated by distorted octahedra. It is well documented that single crystals obtained so far exhibit pronounced disorder between cationic sites. For example, Arima et al [1] reported in their single crystal the site occupancies as follows: Ga1: 0.82Ga, 0.18Fe; Ga2: 0.65Ga, 0.35Fe; Fe1: 0.23Ga, 0.77Fe; Fe2: 0.30Ga, 0.70Fe. Using diffraction and Mössbauer experiments, it was shown that the iron occupancy of Ga tetrahedral sites is much lower than Ga octahedral sites. [1–9]. Thus, one expects that iron in a tetrahedral site is energetically unfavorable. In fact, the hyperfine structure of the $^{57}$Fe nuclear probe at that site is ambiguous because of the low area under the spectra and substantial line overlap. For example, quadrupole splitting for Fe in the Ga1 site was reported to be -0.07 [5] or 0.40(2) [6] (in mm/s).

Stoeffler proposed a model of atomic displacements realizing change between two polarization states of *Pna*2$_1$ structure of fully ordered GaFeO$_3$, $P_z > 0$ (Fig. 1a) and $P_z < 0$ [10] (Fig. 1b). Switching between two polarization states can be realized by a shift of atoms shown schematically by arrows in Fig. 1 a, b. The maximal displacement of oxygen anions deduced from the literature (Table 1) [10] is about 1.2 Å; therefore, it is expected that the switching of the electric polarization cannot be realized easily [11]. Moreover, the switching between two polarization states shown in Fig. 1 [10] changes the local atomic environment. For example, site Fe1 of state $P_z > 0$ changes into site Fe2 of state $P_z < 0$ (and site Fe2 of state $P_z > 0$ changes into site Fe1 of state $P_z < 0$). Similarly, site Ga1 of state $P_z > 0$ changes into site Ga2 of state $P_z < 0$ (and site Ga2 of state $P_z > 0$ changes into site Ga1 of state $P_z < 0$). However, switching between two polarization states proposed by Stoeffler [10] cannot be realized in the case of disordered crystals. As an illustration, let us consider the already mentioned crystal grown by Arima [12]. By switching its polarization state, one would get the partial site occupancies: Ga1: 0.65Ga, 0.35Fe; Ga2: 0.82Ga, 0.18Fe; Fe1: 0.30Ga, 0.70Fe; Fe2: 0.23Ga, 0.77Fe [1]. Thus, the two states ($P_z < 0$ and $P_z > 0$) would differ by atomic disorder and cannot be considered as two opposite polarization states of a ferroelectric. As was already argued, iron in the tetrahedral site is energetically unfavorable; thus, the two states would differ in energy.

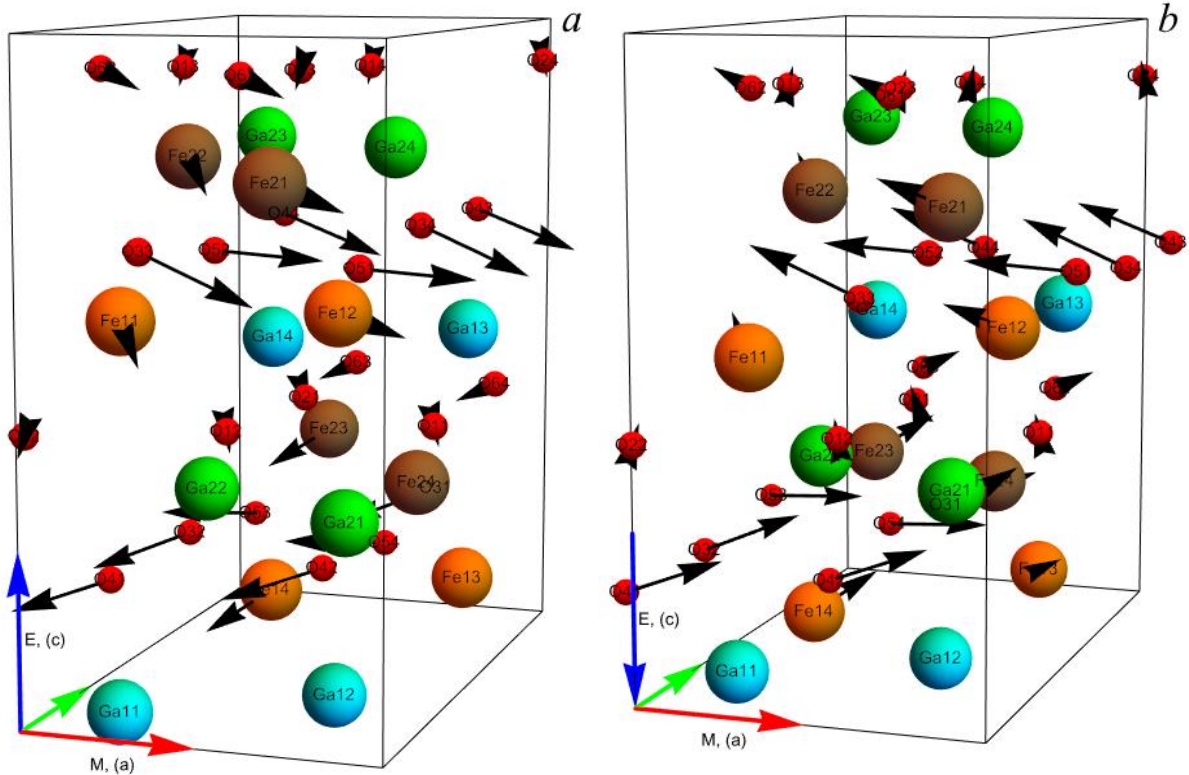

Fig. 1 a) Unit cell in the $P_z > 0$ polarization state. The black arrows show displacements of atoms changing state $P_z > 0$ to the b) state with $P_z < 0$. The black arrows in b) show displacements of atoms changing state $P_z < 0$ to the a) state $P_z > 0$. The view direction is chosen in a way that the largest displacements are clearly shown. The color arrows in the axes origin show directions of magnetization (red, $a$), electrical polarization (blue, $c$), and the third, orthogonal direction (green, $b$). Fe atoms at site 1 related to $Pna2_1$ symmetries: $(x, y, z)$, $(x + 1/2, -y + 1/2, z)$, $(-x, -y, z + 1/2)$ and $(-x + 1/2, y + 1/2, z + 1/2)$ are abbreviated by Fe11, Fe12, Fe13, and Fe14, respectively. The same rule applies to atoms in other sites.

Up to now, the experimental switching between two polarization states was only observed in polycrystalline GaFeO$_3$ [13], in Ga-Fe-O thin films grown by pulsed laser deposition [14–17], on Ga-Fe-O thin films doped by Mg [18], In [19], Cr [20] or Sc [21].

In this paper, we propose an explanation of apparent inconsistency among experimentally observed switching between two polarization states in disordered GaFeO$_3$ crystals and a theoretical description [10] valid only for fully ordered crystals. We provide proof of the electrical polarization switching on a single crystal by piezoresponse atomic force microscopy (PFM). Moreover, we provide electronic structure calculations in particular, the energies related to the effect of disorder. Finally, we compare calculations with the Mössbauer spectroscopy (MS) measurements, providing a consistent description of microscopic GaFeO$_3$ properties.

## 2. Single crystal growth and orientation

The GaFeO$_3$ single crystals were grown by the optical floating zone crystal growth technique according to [1,12] with a four-mirror optical floating zone furnace (FZ-T-4000-H, Crystal Systems Corp. Japan). The starting materials were powders of Fe$_2$O$_3$ (99.999%, Acros Organics) and Ga$_2$O$_3$ (99.99%, Sigma Aldrich) in stoichiometric amounts. The growth was performed in pure oxygen under pressure between 9.0 and 9.2 bar and 0.4 l/min flow rate. Crystals were grown at the rate of 3-5 mm/h, with feed and seed rods rotated at 15 rpm in the opposite directions. The growth direction was enforced using oriented GaFeO$_3$ seeds in $a$ or $c$ crystallographic direction [22,23]. The misorientation of growth directions with respect to the crystalline directions is listed in Table 1.

The crystal orientation was performed using pieces of about 300 μm for which X-ray diffraction data were collected at room temperature. The Rigaku SuperNova diffractometer using Mo Kα radiation ($\lambda$ = 0.71073 Å) was used. Diffraction data were evaluated with the CrysAlisPro package [24]. The crystal structures were solved using direct methods with SHELXT [25] and refined with SHELXL [25] using Independent Atom Model. The GaFeO$_3$ site occupancy was calculated assuming perfect crystal stoichiometry. The results are presented in Table 1.

Table 1. Crystal growth direction misorientation and cation site occupations obtained by single crystal diffraction for the single crystals used in MS and PMF experiments.

| site | site composition Fe/Ga | site composition Fe/Ga |
|---|---|---|
| grown in $a$ direction (2 to 4° misorient., used in MS) | | |
| Fe1 | 0.724(7)/0.276(7) | 0.747(6)/0.253(6) |
| Fe2 | 0.736(8)/0.264(8) | 0.762(7)/0.238(7) |
| Ga1 | 0.081(14)/0.929(14)/ | 0/1 assumed |
| Ga2 | 0.477(10)/0.523(10) | 0.480(10)/0.520(10) |
| grown in $c$ direction (12°misorient., used in MS) | | |
| Fe1 | 0.724(7)/0.276(7) | 0.747(6)/0.253(6) |
| Fe2 | 0.736(8)/0.264(8) | 0.762(7)/0.238(7) |
| Ga1 | 0.074(14)/0.936(14) | 0/1 assumed |
| Ga2 | 0.479(10)/0.521(10) | 0.483(10)/0.517(10) |
| grown in $a$ direction (5°misorient., used in PFM) | | |
| Fe1 | 0.745(8)/0.255(8) | 0.762(7)/0.238(7) |
| Fe2 | 0.733(8)/0.267(8) | 0.748(6)/0.252(6) |
| Ga1 | 0.057(14)/0.953(14) | 0/1 assumed |
| Ga2 | 0.481(10)/0.519(10) | 0.483(10)/0.517(10) |

3. **The microscopic polarity switching mechanism**

Following the available literature [10], a shift of atoms in the GaFeO$_3$ unit cell realizing a polarization switch between two states, $P_z < 0$ and $P_z > 0$, can be found. The condition that the local environments of atoms are preserved, i.e., the local environment of Fe1 in a state with $P_z > 0$ is changing to the same local environment in a state with $P_z < 0$, was also evaluated. A schematic view of the atoms' movements during the polarization switching is shown in Fig. 2. By applying inversion to the structure in Fig. 1a, the $P_z < 0$ unit cell can be obtained (see Table 2, col. 6,7,8), and this structure shifted by vector $s = (0, 0.263, 0.076)$ is shown in Fig. 2b. By the shift $s$, positions of the appropriate pairs of atoms (i.e. those connected by arrows in Fig. 2 a and 2b), in unit cells $P_z > 0$ and $P_z < 0$, are not too far from each other. The most challenging task is to establish correct pairs of atoms. This was done by computer simulations yielding the smallest distance between pairs in the structures $P_z > 0$ and $P_z < 0$. The complete list of pairs involved by the displacement in the unit cell is shown in Table 3.

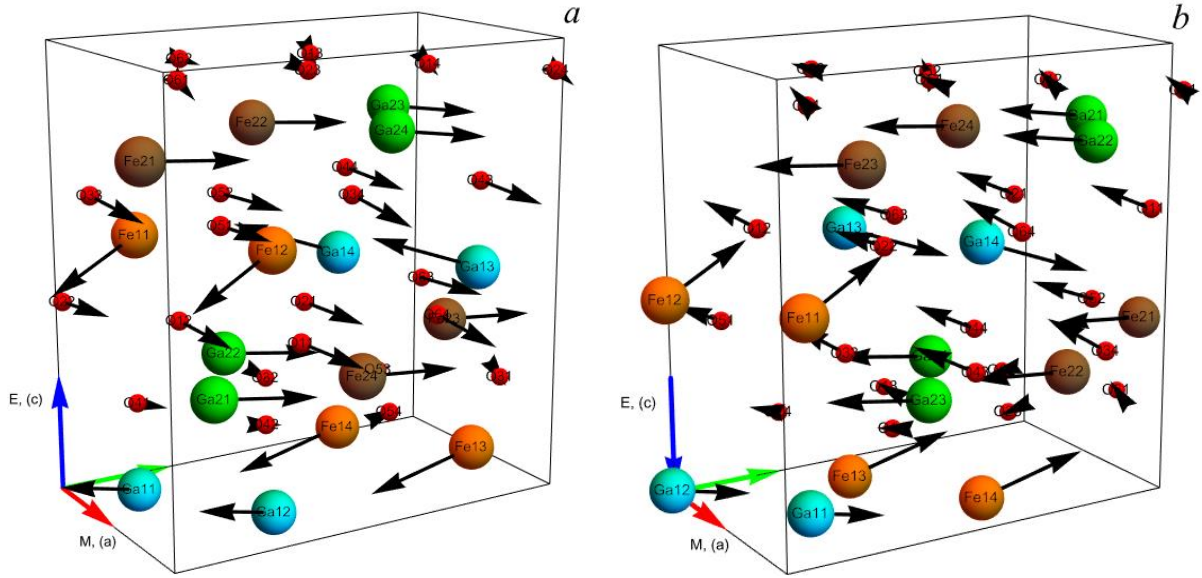

Fig. 2 Schematic view of the proposed switching between two polarization states a) and b) (the description is the same as in the caption of Fig. 1). The structure shown in b) was obtained by inversion of a) by a shift $s = (0, 0.263, 0.076)$.

Table 2. The atomic coordinates for specific polarization states of GaFeO$_3$. Columns 2-4 – $P_z > 0$ state in asymmetric unit; columns 6-8 – $P_z < 0$ state in asymmetric unit, obtained by inversion; columns 9-11 – $P_z < 0$ state shifted by vector $s = (0, 0.263, 0.076)$. The coordinates located outside of the unit cell ($P_z < 0$ state) are shifted by Bravais lattice translations.

| $P_z > 0$ | x | y | z | $P_z < 0$ | x | y | z | x | y | z |
|---|---|---|---|---|---|---|---|---|---|---|
| 1 | 2 | 3 | 4 | 5 | 6 | 7 | 8 | 9 | 10 | 11 |
| Fe11 | 0.1893 | 0.1525 | 0.5827 | Fe12 | 0.3107 | −0.3475 | 0.4173 | 0.3107 | −0.0850 | 0.4937 |
| Fe21 | 0.6787 | 0.0351 | 0.7992 | Fe23 | 0.6787 | 0.0351 | 0.7008 | 0.6787 | 0.2976 | 0.7772 |
| Ga11 | 0.1761 | 0.1501 | 0.0000 | Ga12 | 0.3239 | −0.3499 | 0.0000 | 0.3239 | −0.0874 | 0.0764 |
| Ga21 | 0.8091 | 0.1597 | 0.3067 | Ga23 | 0.8091 | 0.1597 | 0.1933 | 0.8091 | 0.4222 | 0.2697 |
| O11 | 0.9740 | 0.3223 | 0.4260 | O43 | 1.1475 | 0.1593 | 0.3039 | 1.1475 | 0.4218 | 0.3803 |
| O22 | 0.0168 | 0.0123 | 0.4313 | O51 | 0.1590 | −0.1715 | 0.3286 | 0.1590 | 0.0910 | 0.4050 |
| O31 | 0.6521 | 0.9963 | 0.2008 | O61 | 0.4847 | 0.8275 | 0.0621 | 0.4847 | 1.0900 | 0.1385 |
| O42 | 0.6475 | 0.3407 | 0.1961 | O14 | 0.4740 | 0.1777 | 0.0740 | 0.4740 | 0.4402 | 0.1504 |
| O54 | 0.6590 | 0.6715 | 0.1714 | O23 | 0.5168 | 0.4877 | 0.0687 | 0.5168 | 0.7502 | 0.1451 |
| O61 | 0.5153 | 0.1725 | 0.9379 | O31 | 0.3479 | 0.0037 | 0.7992 | 0.3479 | 0.2662 | 0.8756 |

$a$ = 5.0853 Å, $b$ = 8.7451 Å, $c$ = 9.3902 Å

In the proposed switching, the maximal distance followed by cations is 2.3 Å. It is larger than the maximal shift of oxygen anions of 1.2 Å [10]. Nevertheless, the requirement of the same atomic order of both polarization states is fully preserved in the proposed approach.

We have calculated spontaneous polarization as a multivalued vector quantity using point charge approximation and the modern theory of polarization [26]. We get identical values as for the switching proposed in the literature [10]. This is an expected result as the electric polarization can be determined in the Berry phase approach by following a path connecting the polar structure to a centrosymmetric, and the choice of the transition path has no impact on the result itself.

Table 3. Atomic positions in the unit cell of $GaFeO_3$ and details of the atomic displacements corresponding to switching of electrical polarity. Columns 2-4 - atomic coordinates of state $P_z > 0$, columns 6-8 - atomic coordinates of state $P_z < 0$, columns 10-12 - a shift of atoms between $P_z > 0$ and $P_z < 0$ states. Column 9 shows the displacement of atoms when switching between $P_z > 0$ and $P_z < 0$.

|   | site $P_z >0$ | x | y | z | site $P_z <0$ | x | y | z | d [Å] | $d_x$ | $d_y$ | $d_z$ |
|---|---|---|---|---|---|---|---|---|---|---|---|---|
|   | 1 | 2 | 3 | 4 | 5 | 6 | 7 | 8 | 9 | 10 | 11 | 12 |
| 1 | Fe11 | 0.1893 | 0.1525 | 0.5827 | Fe12 | 0.3107 | −0.0850 | 0.4937 | 2.3220 | 0.2428 | −0.4750 | −0.1780 |
| 2 | Fe12 | 0.6893 | 0.3475 | 0.5827 | Fe11 | 0.8107 | 0.1100 | 0.4937 | 2.3220 | 0.2428 | −0.4750 | −0.1780 |
| 3 | Fe13 | 0.8107 | 0.8475 | 0.0827 | Fe14 | 0.6893 | 0.6100 | −0.0063 | 2.3219 | −0.2428 | −0.4750 | −0.1780 |
| 4 | Fe14 | 0.3107 | 0.6525 | 0.0827 | Fe13 | 0.1893 | 0.4150 | −0.0063 | 2.3219 | −0.2428 | −0.4750 | −0.1780 |
| 5 | Fe21 | 0.6787 | 0.0351 | 0.7992 | Fe23 | 0.6787 | 0.2976 | 0.7772 | 2.3052 | 0.0000 | 0.5250 | −0.0440 |
| 6 | Fe22 | 0.1787 | 0.4649 | 0.7992 | Fe24 | 0.1787 | 0.7274 | 0.7772 | 2.3052 | 0.0000 | 0.5250 | −0.0440 |
| 7 | Fe23 | 0.3213 | 0.9649 | 0.2992 | Fe21 | 0.3213 | 1.2274 | 0.2772 | 2.3052 | 0.0000 | 0.5250 | −0.0440 |
| 8 | Fe24 | 0.8213 | 0.5351 | 0.2992 | Fe22 | 0.8213 | 0.7976 | 0.2772 | 2.3052 | 0.0000 | 0.5250 | −0.044 |
| 9 | Ga11 | 0.1761 | 0.1501 | 0.0000 | Ga12 | 0.3239 | −0.0874 | 0.0764 | 2.3221 | 0.2956 | −0.4750 | 0.1528 |
| 10 | Ga12 | 0.6761 | 0.3499 | 0.0000 | Ga11 | 0.8239 | 0.1124 | 0.0764 | 2.3221 | 0.2956 | −0.4750 | 0.1528 |
| 11 | Ga13 | 0.8239 | 0.8499 | 0.5000 | Ga14 | 0.6761 | 0.6124 | 0.5764 | 2.3220 | −0.2956 | −0.4750 | 0.1528 |
| 12 | Ga14 | 0.3239 | 0.6501 | 0.5000 | Ga13 | 0.1761 | 0.4126 | 0.5764 | 2.3220 | −0.2956 | −0.4750 | 0.1528 |
| 13 | Ga21 | 0.8091 | 0.1597 | 0.3067 | Ga23 | 0.8091 | 0.4222 | 0.2697 | 2.3221 | 0.0000 | 0.5250 | −0.0740 |
| 14 | Ga22 | 0.3091 | 0.3403 | 0.3067 | Ga24 | 0.3091 | 0.6028 | 0.2697 | 2.3221 | 0.0000 | 0.5250 | −0.0740 |
| 15 | Ga23 | 0.1909 | 0.8403 | 0.8067 | Ga21 | 0.1909 | 1.1028 | 0.7697 | 2.3221 | 0.0000 | 0.5250 | −0.0740 |
| 16 | Ga24 | 0.6909 | 0.6597 | 0.8067 | Ga22 | 0.6909 | 0.9222 | 0.7697 | 2.3221 | 0.0000 | 0.5250 | −0.0740 |
| 17 | O11 | 0.9740 | 0.3223 | 0.4260 | O43 | 1.1475 | 0.4218 | 0.3803 | 1.3117 | 0.3470 | 0.1990 | −0.0914 |
| 18 | O12 | 0.4740 | 0.1777 | 0.4260 | O33 | 0.6521 | 0.2588 | 0.3756 | 1.2442 | 0.3562 | 0.1622 | −0.1008 |
| 19 | O13 | 0.0260 | 0.6777 | 0.9260 | O32 | −0.1521 | 0.7588 | 0.8756 | 1.2439 | −0.3562 | 0.1622 | −0.1008 |
| 20 | O14 | 0.5260 | 0.8223 | 0.9260 | O42 | 0.3525 | 0.9218 | 0.8803 | 1.3115 | −0.3470 | 0.1990 | −0.0914 |
| 21 | O21 | 0.5168 | 0.4877 | 0.4313 | O44 | 0.6475 | 0.6032 | 0.3803 | 1.3009 | 0.2614 | 0.2310 | −0.1020 |
| 22 | O22 | 0.0168 | 0.0123 | 0.4313 | O51 | 0.1590 | 0.0910 | 0.4050 | 1.0287 | 0.2844 | 0.1574 | −0.0526 |
| 23 | O23 | 0.4832 | 0.5123 | 0.9313 | O54 | 0.3410 | 0.5910 | 0.9050 | 1.0285 | −0.2844 | 0.1574 | −0.0526 |
| 24 | O24 | 0.9832 | 0.9877 | 0.9313 | O41 | 0.8525 | 1.1032 | 0.8803 | 1.3007 | −0.2614 | 0.2310 | −0.1020 |
| 25 | O31 | 0.6521 | 0.9963 | 0.2008 | O61 | 0.4847 | 1.0900 | 0.1385 | 1.3185 | −0.3348 | 0.1874 | −0.1246 |
| 26 | O32 | 0.1521 | 0.5037 | 0.2008 | O13 | −0.0260 | 0.5848 | 0.1504 | 1.2439 | −0.3562 | 0.1622 | −0.1008 |
| 27 | O33 | 0.3479 | 0.0037 | 0.7008 | O12 | 0.5260 | 0.0848 | 0.6504 | 1.2442 | 0.3562 | 0.1622 | −0.1008 |
| 28 | O34 | 0.8479 | 0.4963 | 0.7008 | O64 | 1.0153 | 0.5900 | 0.6385 | 1.3188 | 0.3348 | 0.1874 | −0.1246 |
| 29 | O41 | 0.1475 | 0.1593 | 0.1961 | O24 | 0.0168 | 0.2748 | 0.1451 | 1.3007 | −0.2614 | 0.2310 | −0.1020 |
| 30 | O42 | 0.6475 | 0.3407 | 0.1961 | O14 | 0.4740 | 0.4402 | 0.1504 | 1.3115 | −0.3470 | 0.1990 | −0.0914 |
| 31 | O43 | 0.8525 | 0.8407 | 0.6961 | O11 | 1.0260 | 0.9402 | 0.6504 | 1.3117 | 0.3470 | 0.1990 | −0.0914 |
| 32 | O44 | 0.3525 | 0.6593 | 0.6961 | O21 | 0.4832 | 0.7748 | 0.6451 | 1.3009 | 0.2614 | 0.2310 | −0.1020 |
| 33 | O51 | 0.8410 | 0.1715 | 0.6714 | O22 | 0.9832 | 0.2502 | 0.6451 | 1.0287 | 0.2844 | 0.1574 | −0.0526 |
| 34 | O52 | 0.3410 | 0.3285 | 0.6714 | O63 | 0.5153 | 0.4350 | 0.6385 | 1.3227 | 0.3486 | 0.2130 | −0.0658 |
| 35 | O53 | 0.1590 | 0.8285 | 0.1714 | O62 | −0.0153 | 0.9350 | 0.1385 | 1.3224 | −0.3486 | 0.2130 | −0.0658 |
| 36 | O54 | 0.6590 | 0.6715 | 0.1714 | O23 | 0.5168 | 0.7502 | 0.1451 | 1.0285 | −0.2844 | 0.1574 | −0.0526 |
| 37 | O61 | 0.5153 | 0.1725 | 0.9379 | O31 | 0.3479 | 0.2662 | 0.8756 | 1.3185 | −0.3348 | 0.1874 | −0.1246 |
| 38 | O62 | 0.0153 | 0.3275 | 0.9379 | O53 | −0.1590 | 0.4340 | 0.9050 | 1.3224 | −0.3486 | 0.2130 | −0.0658 |
| 39 | O63 | 0.4847 | 0.8275 | 0.4379 | O52 | 0.6590 | 0.9340 | 0.4050 | 1.3227 | 0.3486 | 0.2130 | −0.0658 |
| 40 | O64 | 0.9847 | 0.6725 | 0.4379 | O34 | 1.1521 | 0.7662 | 0.3756 | 1.3188 | 0.3348 | 0.1874 | −0.1246 |

## 4. Details of ab initio calculations

The first-principles calculations were performed using the projector augmented wave (PAW) potentials [27] and the generalized gradient approximation (GGA) in the Pardew, Burke, and Ernzerhof (PBE) parametrization [28]. We have used the VASP code [29–31], and the calculations were based on the stoichiometric GaFeO$_3$. We included three valence electrons for Ga atoms ($4s^2\ 4p^1$), eight for Fe atoms ($3d^7\ 4s^1$), and six for O atoms ($2s^2\ 2p^4$). The Hubbard parameter $U$ and exchange interaction $J$ were optimized for the ion's magnetic moment to fit with the experiment ($U = 8$ eV, $J = 1$ eV) [1,32,33]. A plane wave energy cutoff of 520 eV was used. The conjugate gradient algorithm was used to optimize the structure with the energy convergence criteria set at $10^{-8}$ and $10^{-5}$ eV for electronic and ionic iterations, respectively. For the summation over the reciprocal space, we used a 10 x 6 x 6 Monkhorst-Pack $k$-point grid [34]. The simulations were performed for the orthorhombic unit cell, comparing eight formula units (40 atoms). The electric field gradient tensors at the positions of the atomic nuclei are calculated using the method described in the literature [35,36]. The results are presented in Table 4.

Table 4. Theoretical values of the EFG tensor components $V_{ij}$ of $^{57}$Fe probe at cationic sites of GaFeO$_3$. Symbol $V_{kk}$ is the dominant component of the EFG tensor in the local principal axes system $e_i$, $e_j$, $e_k$, where their Cartesian components in the unit cell frame $x$, $y$, $z$, are listed in columns 4 to 12; indices $i,j,k$ are ordered so that $|V_{ii}| \leq |V_{jj}| \leq |V_{kk}|$. The asymmetry parameter (col. 3) is defined as $\eta = (V_{ii} - V_{jj})/V_{kk}$. The superscript star in the first column indicates the results of calculations for single iron at the foreign site of GaFeO$_3$.

|  | $V_{kk}$ [V/Å$^2$] | $\eta$ | $e_i$ | | | $e_j$ | | | $e_k$ | | |
|---|---|---|---|---|---|---|---|---|---|---|---|
|  |  |  | $e_i \cdot x$ | $e_i \cdot y$ | $e_i \cdot z$ | $e_j \cdot x$ | $e_j \cdot y$ | $e_j \cdot z$ | $e_k \cdot x$ | $e_k \cdot y$ | $e_k \cdot z$ |
| 1 | 2 | 3 | 4 | 5 | 6 | 7 | 8 | 9 | 10 | 11 | 12 |
| Fe11 | −31.989 | 0.627 | 0.768 | −0.516 | −0.379 | −0.075 | 0.516 | −0.853 | 0.636 | 0.684 | 0.358 |
| Fe21 | 34.138 | 0.995 | −0.287 | 0.172 | 0.942 | 0.956 | 0.113 | 0.271 | −0.060 | 0.978 | −0.198 |
| Ga11 | −30.032 | 0.413 | −0.395 | 0.811 | 0.432 | 0.253 | −0.356 | 0.900 | 0.883 | 0.464 | −0.064 |
| Ga21 | −56.012 | 0.460 | 0.825 | −0.554 | 0.112 | 0.353 | 0.350 | −0.868 | 0.442 | 0.756 | 0.484 |
| Ga11* | −32.885 | 0.178 | −0.457 | 0.773 | 0.440 | 0.262 | −0.356 | 0.897 | 0.850 | 0.525 | −0.040 |
| Ga21* | −71.963 | 0.485 | 0.822 | −0.556 | 0.121 | 0.359 | 0.342 | −0.868 | 0.442 | 0.757 | 0.481 |

Table 5. Theoretical values of the parameters of the Mössbauer spectra. $QS = \frac{eQc|V_{zz}|}{2E_0}\sqrt{1+\eta^2/3}$ - separation between two absorption lines in the quadrupole doublet, $A_1/A_2$ - a ratio of the absorption line intensities in the doublet, $k$ - direction of the wave vector of a photon with respect to the main crystal direction (reference of *Pna2$_1$* space group). Superscript *asterisk* (*) in the first column indicates the results of calculations for single iron at the foreign site of GaFeO$_3$. Symbol $e$ is the elementary charge, $Q$ nuclear quadrupole moment of the first excited state of $^{57}$Fe ($1.7 \cdot 10^{-29}$ m$^2$), $E_0$ is the energy of a photon in Mössbauer excitation (14.412497 keV), $c$ - speed of light.

| $k$ | 100 | 100 | 100 | 100 | 010 | 010 | 010 | 010 | 001 | 001 | 001 | 001 |
|---|---|---|---|---|---|---|---|---|---|---|---|---|
| site | Fe1 | Fe2 | Ga1 | Ga2 | Fe1 | Fe2 | Ga1 | Ga2 | Fe1 | Fe2 | Ga1 | Ga2 |
| 1 | 2 | 3 | 4 | 5 | 6 | 7 | 8 | 9 | 10 | 11 | 12 | 13 |
| $QS$ | 0.30 | 0.35 | 0.29 | 0.66 | 0.30 | 0.35 | 0.29 | 0.66 | 0.30 | 0.35 | 0.29 | 0.66 |
| $A_1/A_2$ | 1.31 | 2.3 | 1.84 | 0.93 | 1.21 | 0.42 | 0.96 | 1.49 | 0.62 | 1.03 | 0.57 | 0.72 |
| site | Fe1 | Fe2 | Ga1* | Ga2* | Fe1 | Fe2 | Ga1* | Ga2* | Fe1 | Fe2 | Ga1* | Ga2* |
| $QS$ | 0.3 | 0.35 | 0.27 | 0.51 | 0.3 | 0.35 | 0.27 | 0.51 | 0.3 | 0.35 | 0.27 | 0.51 |
| $A_1/A_2$ | 1.31 | 2.3 | 2.01 | 0.93 | 1.21 | 0.42 | 0.94 | 1.48 | 0.62 | 1.03 | 0.54 | 0.73 |

The first principle calculations show evident change of the EFG when Fe impurity enters into Ga sites. This effect could not be obtained in point-charge calculations performed earlier [8,37]. Moreover, the $QS$ splitting of Ga1 is very close to that of Fe1 (compare $QS$ in col. 2 and 4, 6 and 8, 10 and 12 in Table 5). This may be the reason why only three components were detected in Mössbauer experiments reported so far.

## 5. Results of atomic disorder energy calculations

To determine, which positions are energetically favorable for additional Fe atoms, we have calculated crystal energy when one Ge atom is substituted by a Fe atom in the unit cell. The calculations were done for two non-equivalent Ga sites, resulting in the off-stoichiometric crystal $Ga_{2-x}Fe_xO_3$ with x=1.125, see Table 6. The additional Fe atom prefers to occupy the Ga2 site (with the magnetic moment direction at the additional Fe atom the same as for Fe1 atoms). On the other hand, the case of one Fe atom substituted by a Ga atom in two non-equivalent Fe sites ($Ga_{2-x}Fe_xO_3$ with x=0.875) leads to similar energies. Thus, both additional Fe atom positions are almost equally probable, while additional Ga goes to either Fe1 or Fe2 site, see Table 6.

Table 6. Energies of nonstoichiometric $Ga_{2-x}Fe_xO_3$ crystal with a different type of disorder. In the unit cell composed of 40 atoms, one Fe atom was substituted by one Ga atom, or one Ga atom was substituted by one Fe atom. The energy of the ideal crystal is $E_0 = -250.6123346$ eV.

|    | x     | site disorder | energy [eV]   |
|----|-------|---------------|---------------|
| 1. | 1.125 | Fe in Ga1     | −251.632745   |
| 2. | 1.125 | Fe in Ga2     | −251.8857439  |
| 3. | 0.875 | Ga in Fe1     | −249.2106457  |
| 4. | 0.875 | Ga in Fe2     | −249.2295923  |

We also investigated the effect of atomic disorder in the form of an interchange between Fe and Ga sites, as shown in Table 7. For example, for a given Fe1 position, there are four positions of Ga1 in the unit cell, corresponding to four symmetry operations of the *Pna*$2_1$ structure, abbreviated by Ga11, Ga12, Ga13, and Ga14, respectively. Let us underline that all calculations presented in this section were performed for structures (atom locations) as in the optimized cell of the ideal crystal.

Table 7. The energy of site disorder. A pair of Fe and Ga atoms (column 2) interchange their positions in the unit cell resulting in energy listed in column 3. Values averaged over sites generated by symmetry operations are listed in column 4.

|   | interchange of two atoms between sites | energy $E$ | $\bar{E} - E_0$ |
|---|---|---|---|
| 1 | 2 | 3 | 4 |
| 1. | Fe11, Ga11 | −250.2321563 | |
| 2. | Fe11, Ga12 | −250.2413128 | 0.378 |
| 3. | Fe11, Ga13 | −250.2413128 | |
| 4. | Fe11, Ga14 | −250.2329995 | |
| 5. | Fe11, Ga21 | −250.4597895 | |
| 6. | Fe11, Ga22 | −250.4776731 | 0.143 |
| 7. | Fe11, Ga23 | −250.4564648 | |
| 8. | Fe11, Ga24 | −250.4833769 | |
| 9. | Fe21, Ga11 | −250.2065841 | |
| 10. | Fe21, Ga12 | −250.2365022 | 0.384 |
| 11. | Fe21, Ga13 | −250.2498087 | |
| 12. | Fe21, Ga14 | −250.2225650 | |
| 13. | Fe21, Ga21 | −250.5050056 | |
| 14. | Fe21, Ga22 | −250.5045072 | 0.112 |
| 15. | Fe21, Ga23 | −250.4888525 | |
| 16. | Fe21, Ga24 | −250.5055941 | |

### 6. Mössbauer experiments

Details of the Mössbauer experiment are given in [38]. The single crystals grown in $a$ and in $c$ directions (Table 1) were used for preparation single crystalline, oriented absorbers. Using electronic structure calculations on ideal GaFeO$_3$ and on single Fe cations at Ga sites, we obtained electric field gradients on [57]Fe probe at cationic sites (Table 4). With the help of the full Hamiltonian formalism [37] adapted to GaFeO$_3$ crystal [22], we have calculated parameters of doublets for [57]Fe iron probe at different sites and orientations of $k$ vector (Table 5).

The experimental spectra of GaFeO$_3$ consist of overlapping subspectra, and the assignment of absorption lines to the sites is not apparent. Also, the ambiguity problem is present, i.e., the continuous distribution of parameters results in the exact shape of the spectrum. The measurements of texture-free absorbers in external magnetic fields were performed to make the interpretation precise. Moreover, the measurements of single crystal absorbers with the orientation of wave vector $k$ parallel to the main crystal directions were done. One of the most difficult problems is spectra interpretation; the orientation of the EFG was solved by adopting principal directions obtained by electronic structure calculations. Also, the asymmetry parameter in the in-magnetic field experiment was adopted from the theoretical calculations. It is worthwhile to add that throughout the paper, we use four colors related to the four cationic sites, already shown on unit cells (Fig. 1 and 2) and also used for the abbreviation of the subspectra in Fig. 3 and 4 or sites in Fig. 7 and 8.

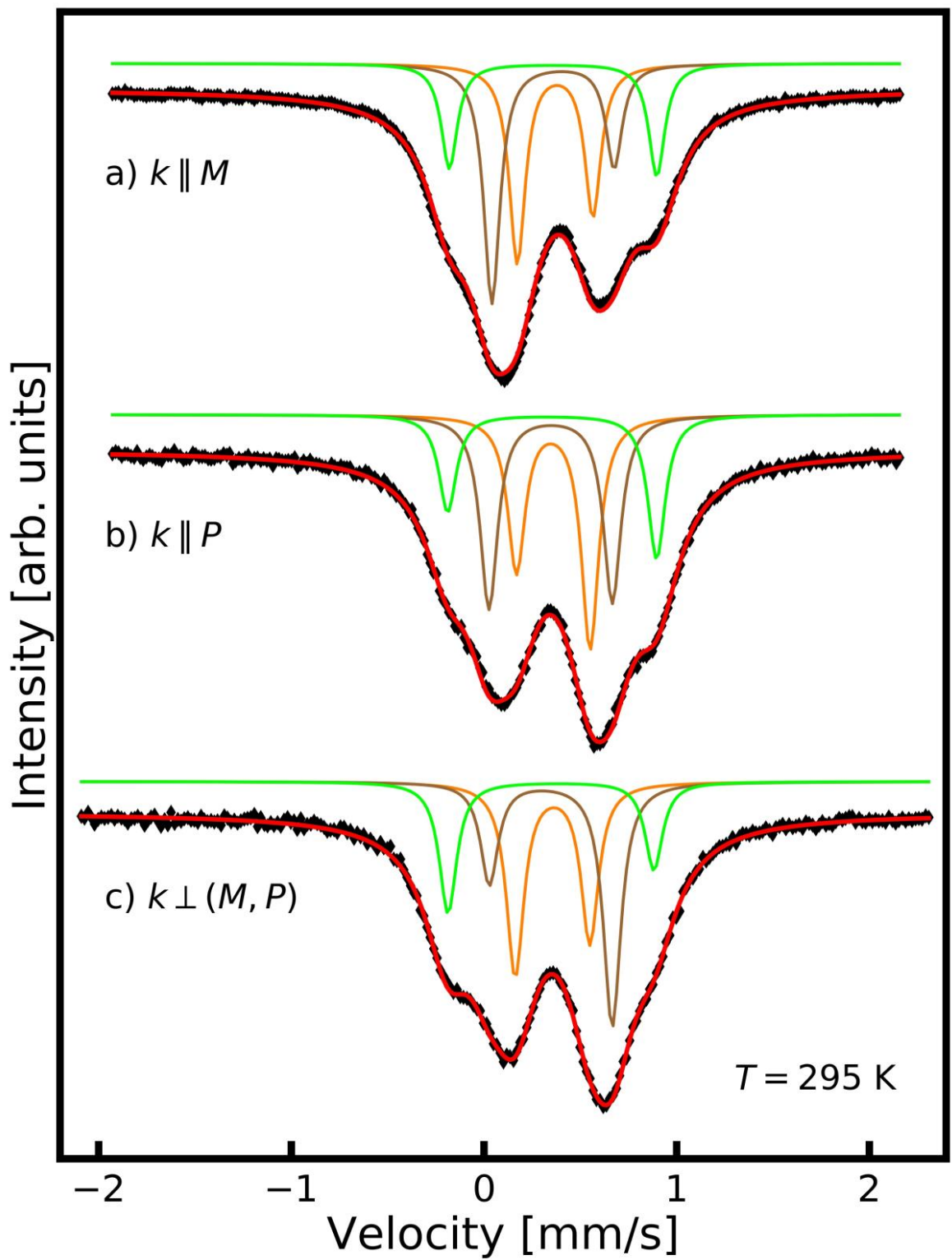

Fig. 3 Mössbauer spectra of single crystals measured with $k$ - vector parallel to the three main crystallographic directions. Relative line intensities in the doublets are taken from theoretical calculations (Table 4), while other parameters are listed in Table 5. Color curves represent subspectra of iron located in Fe1 (orange), Fe2 (brown), and Ga2 green) sites. The red curve is a fit to experimental data (black points).

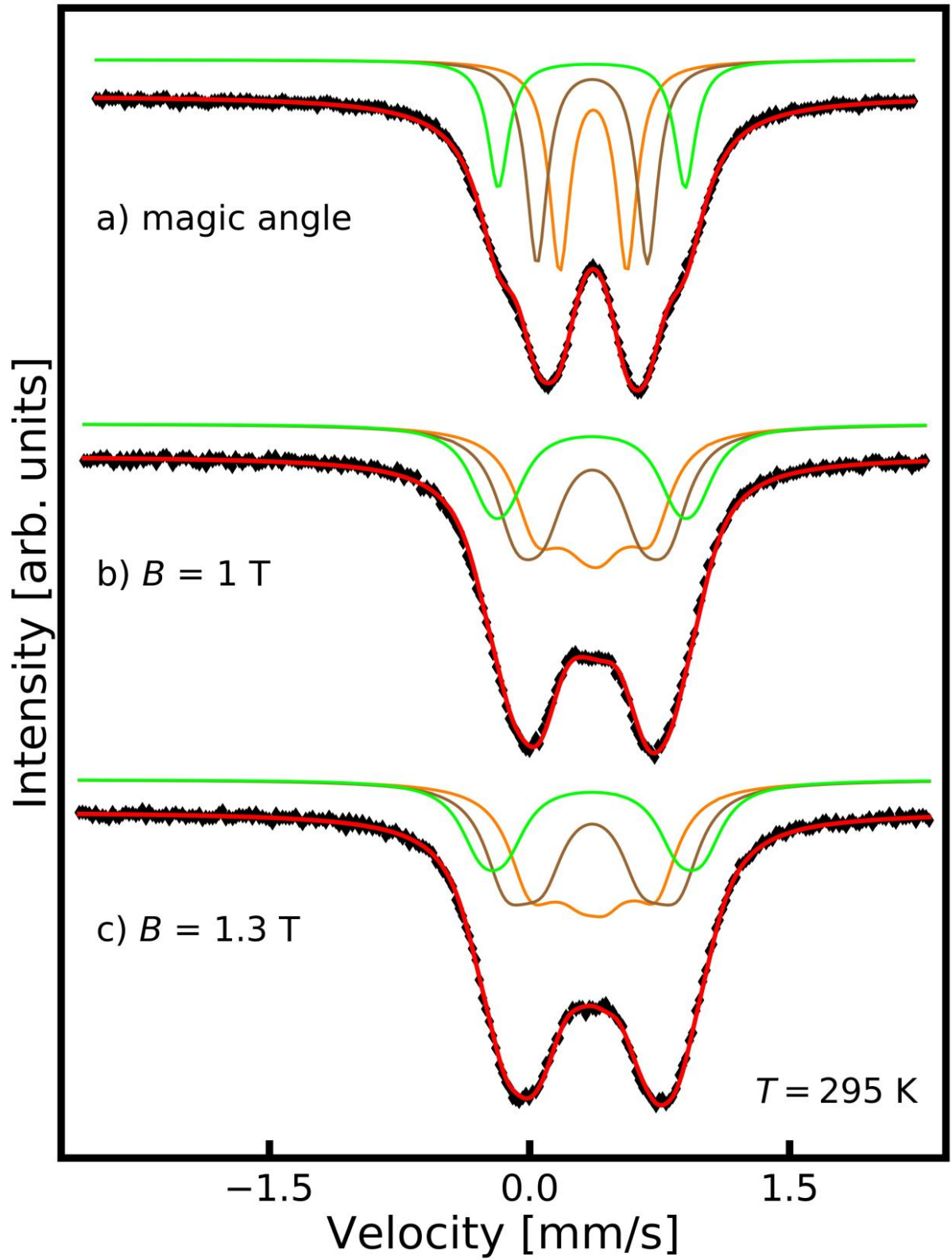

Fig. 4 Mössbauer spectra of powdered single crystals measured at a magic angle a) and in external magnetic fields b), c). Relative line intensities in the doublets are taken from theoretical calculations (Table 4). The description of color curves is the same as in Fig. 3.

The results of single crystals measurements, with $k$ - vector parallel to the three main crystallographic directions, are presented in Fig. 3. Relative line intensities in the doublets were taken from theoretical calculations [37] (Table 5, $A_1/A_2$). The results from the powdered single crystals are shown in Fig. 4. In order to avoid the problem with the crystal texture, the measurements with a so-called magic angle were done [38]. The measured spectrum (see Fig. 4a) is a superposition of a few not well-resolved doublets from the iron atoms in the GaFeO$_3$ structure. In order to get more data, the measurements in an external magnetic field of $B$ = 1 T and 1.3 T, parallel to the beam direction, were conducted. In the analysis of in-field measurements, the Blaes procedure was used [39]. The hyperfine fields induced in the local position in the GaFeO$_3$ were free parameters in the fitting. The simultaneous fit shows that three subspectra are needed to describe the shapes of measured spectra well. The solid red line represents the best fit, while the other lines represent three subspectra. The hyperfine parameters and relative intensities of the doublets are presented in Table 8. The analysis shows that Fe atoms occupy only three octahedral positions in the crystal structure of GaFeO$_3$. Almost the same fraction of iron occupies the Fe1 and Fe1 sites, while in Ga2 sites, the fraction of iron is twice smaller than in Fe1 and Fe2 sites. There is no experimental evidence within the limits of experimental uncertainty that Fe is located in the tetrahedral Ga sites.

Table 8. Parameters of hyperfine interactions for subspectra are shown in Fig. 3 and 4. The asymmetry parameter $\eta$ was adopted from theoretical calculations (Table 4). Sub1, Sub2 Sub3 are spectral areas proportional to the number of Fe atoms in the site; Sub4=0 was assumed since this spectral area was below a detection limit.

| Sample | *ISO*±0.02 [mm/s] | | | *QS*±0.02 [mm/s] | | | *BHF*±0.07 [T] | | | *ARE*±2 [%] | | |
|---|---|---|---|---|---|---|---|---|---|---|---|---|
| | Sub1 | Sub2 | Sub3 | Sub1 | Sub2 | Sub3 | Sub1 | Sub2 | Sub3 | Sub1 | Sub2 | Sub3 |
| single crystal | 0.36 | 0.35 | 0.36 | −0.39 | 0.64 | −1.08 | --- | --- | --- | 38 | 38 | 24 |
| powder, magic angle | 0.37 | 0.36 | 0.36 | −0.39 | 0.64 | −1.08 | --- | --- | --- | 38 | 38 | 24 |
| powder, $B_{ext}$ = 1 T | 0.35 | 0.35 | 0.34 | −0.36 | 0.62 | −1.05 | 1.16 | 0.89 | 0.53 | 38 | 38 | 24 |
| powder, $B_{ext}$ = 1.3 T | 0.36 | 0.35 | 0.35 | −0.40 | 0.68 | −1.10 | 1.22 | 0.95 | 0.59 | 38 | 38 | 24 |

## 7. Piezoresponse Force Microscopy

Piezoresponse force microscopy (PFM) is designed to measure piezoelectric and ferroelectric materials on the nanoscale [40]. All measurements used NanoWizard®3 BioScience (JPK Instruments, Berlin, Germany) AFM. Images were acquired using conductive diamond-coated AFM tips with resonant frequencies close to 110 kHz. The experimental setup allowed tip/detector calibration to measure PFM response in pm. The modulation voltage was set to 1 V, which gave a relatively strong PFM signal. In addition, the bias voltage was applied in the range of 0 to 20 V, depending on the experiment. The piezo-switching (images) were obtained with no bias, where the polarization change was forced by +5 V bias, while the PFM spectroscopy was performed in the range of ± 20 V. The data analysis was performed using Gwyddion software [41].

The sample plane of dimension 3.9 x 3.9 mm was cut from a single crystal grown along $a$ axis ($Pna2_1$ space group). The plane was perpendicular to the $c$ crystallographic axis within the accuracy of 2°. By intensive PFM investigations, we could not find visible ferroelectric domain structure registered down to the nanoscale range. It suggests that the sample was a single domain or that the domain size was above the maximum possible scan size of implemented AFM equipment, which was 100 x 100 µm. However, to the best of our knowledge, no such ferroelectric domains (on single crystals) were found using other techniques like surface decoration, etching, optical microscopy, polarized light microscopy, X-ray techniques, or electron microscopy techniques. It is consistent with the fact that relatively large energy is needed to switch between opposite polarization, as mentioned previously.

Nevertheless, the piezoresponse signal from the sample was relatively strong, showing typical piezoelectric behavior. To manifest that, two types of experiments were performed. First, we switched the material's polarization by applying a bias voltage to a small region (in this case, 5 x 5 µm and +5 V bias voltage). We measured the same region in a larger scan (10x10 µm, no extra bias voltage). The result is shown in Fig. 5, where the PFM amplitude and phase are presented. The data is presented as a pseudo-3D map, where the value is correlated with the color. Thus, the higher the color – the higher the value. One can see that it is possible to change the polarization of the $GaFeO_3$ single crystal. However, due to the small contact area between the conducting AFM tip and the sample surface, the electric field applied in the PFM technique was relatively large, the range of 500 kV/cm$^{-1}$ (assuming 100 nm size of tip-sample contact).

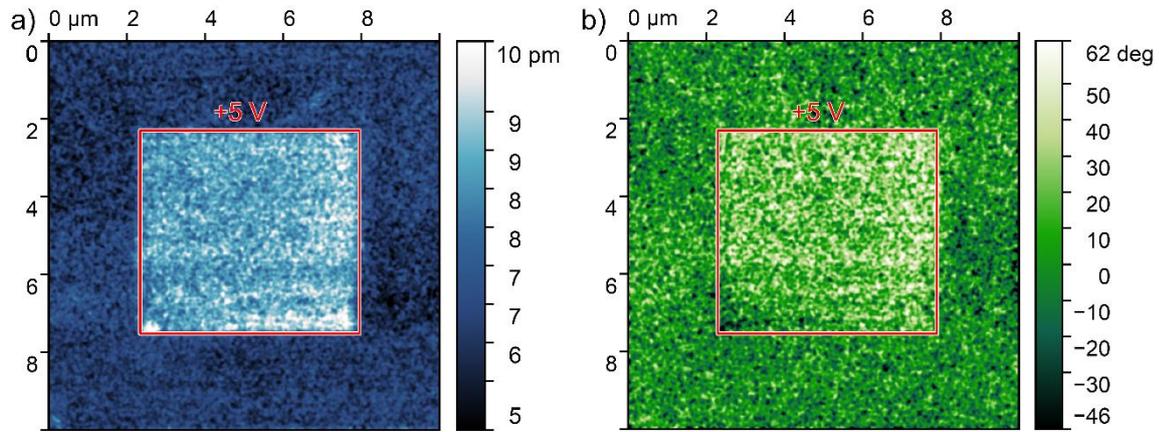

Fig. 5 PFM images of oriented GaFeO$_3$ single crystal showing a) amplitude and b) phase-contrast images acquired over 10x10 μm area. The images are a result of scanning the central region (5 x 5 μm area) with the positive bias of +5 V, subsequently switching the piezoelectric polarization of the region.

Next, a PFM spectroscopy in a single point was performed in the range of ±20 V. The results are shown in Fig. 6. The PFM amplitude and phase contrast are shown as a function of bias voltage. The switching between the two polarization states is visible, with the coercive voltage close to 2 V. The phase change (Fig.6 b) also shows the change in the polarization state of 180°, further confirming that the switching occurred between two opposite polarization states.

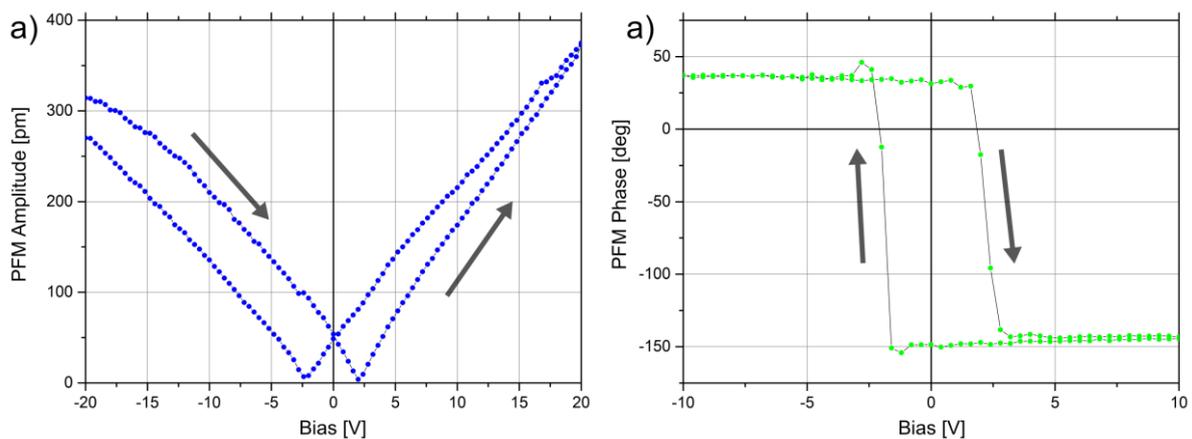

Fig.6, PFM spectroscopy curves of oriented GaFeO$_3$ single crystal showing a) amplitude and b) phase-change signals as a function of bias voltage performed at a random sample position. The results show

switching between two piezoelectric polarization states. Arrows donate the direction of spectroscopy: the voltage changed from – 20 V to +20 V and back.

## 8. Discussion

The results of theoretical calculations of the electric field gradient tensor allow the use of its local orientations in data analysis. The relative line intensities of Mössbauer spectra shown in Fig. 3 were calculated without any fitting procedure. Also, the predicted asymmetry parameter (Table 4) was used in the data analysis of spectra measured in an external magnetic field (Fig. 4). These theoretical predictions agree perfectly with the experimental observations (Fig. 3). The predicted values of EFG tensor components (Table 5, row 2, $QS$) are, however, systematically smaller than observed values (Table 8) by about a factor of 2.

The atomic disorder measured by X-ray diffraction (Table 1) is consistent with that determined in Mössbauer experiments (Table 9). It plays a crucial role in our investigations, as it allows the estimation of the atoms' energies entering the foreign sites. It can also serve as the evaluation for the model predicting the electrical switching polarization.

It is clear from Table 6 (rows 1,2) that Fe in the Ga1 site has larger energy than in the Ga2 site by about 0.257 eV. This value coincides nicely with energies of interchange presented in Table 6 for defect creation of Fe in the Ga1 position. This energy is 0.384 eV and 0.378 eV on average: Fe2-Ga1 slightly larger than Fe1-Ga1. The calculations also show that the energy of Ga location in foreign sites Fe1 and Fe2 are very similar. In crystals with an excess of gallium x = 0.875 (Table 6, rows 3,4), Ga in the Fe1 site has larger energy than in the Fe2 site by 0.025 eV. Analysis of pair interchange in stoichiometric crystal shows a similar effect; the interchange of Fe1 and Ga2 atoms requires energy larger by about 0.031 eV than that of Fe2 and Ga2. One expects Ga in both Fe2 and Fe2 sites to locate easily, while Fe should hardly enter Ga1 sites.

The results of theoretical calculations are consistent with a simple model of structural defects. Let us simplify interactions by assuming that, on average, Fe or Ga atoms entering the foreign site increase the crystal energy by a value independent of their geometrical arrangement in the unit cell. This assumption is governed by the observation in Table 7 that in rows 1 to 4, the energies are similar to each other, and in rows 5 to 8, 9 to 12, and 13 to 16 as well. We also neglect the compositional dependence of the energies in the range covered in Table 6. We thus introduce energy $E_{\text{Fe}}^{(1)}$ corresponding to the presence of foreign Fe in the Ga1 site. Further energies are defined consequently: $E_{Fe}^{(2)}$, $E_{Ga}^{(1)}$ and $E_{Ga}^{(2)}$. We may write for the third column of Table 6:

$$E_{Fe}^{(2)} - E_{Fe}^{(1)} - u_2 + u_1 = 0, \quad (1)$$

$$E_{Ga}^{(2)} - E_{Ga}^{(1)} - u_4 + u_3 = 0, \quad (2)$$

$$E_{Ga}^{(1)} + E_{Ga}^{(2)} + E_{Fe}^{(1)} + E_{Fe}^{(2)} - 4e_0 = 0,$$

where $u_i$ is given in the i-the row of Table 5 and $e_0$ is the energy of an unperturbed crystal, shown in the caption of Table 6. Similarly

$$E_{Fe}^{(1)} + E_{Ga}^{(1)} - v_{1,4} + e_0 = 0, \quad (3)$$

$$E_{Fe}^{(2)} + E_{Ga}^{(1)} - v_{5,8} + e_0 = 0, \quad (4)$$

$$E_{Fe}^{(1)} + E_{Ga}^{(2)} - v_{9,12} + e_0 = 0, \quad (5)$$

$$E_{Fe}^{(2)} + E_{Ga}^{(2)} - v_{13,16} + e_0 = 0, \quad (6)$$

where $v_{i,j}$ is given by any row between $i$ and $j$ of Table 7. Equations (1) to (6) are mathematically contradictory. However, an approximation can be obtained if energies $E_{Fe}^{(1)}$, $E_{Fe}^{(2)}$, $E_{Ga}^{(1)}$ and $E_{Ga}^{(2)}$ are chosen so that sets (1) to (6) are fulfilled approximately, with possibly minor deviations. To find the energies, we minimize the sum of squares of the left-hand side of eq. (1) to (6) with additional physical constraints for the energies to be positive. There are $4 \cdot 4 \cdot 4 \cdot 4$ numbers of choices of the energies $v_{1,4}$, $v_{5,8}$, $v_{9,12}$ and $v_{13,16}$ taken from Table 7, which appears in eq. (3) to (6). Calculations show that for all the choices, calculated energies are distributed within a relatively narrow range of values, see the histogram in Fig. 7. The average values and standard deviations are equal to: $E_{Fe}^{(1)} = 0.299(5)$, $E_{Fe}^{(2)} = 0.046(2)$, $E_{Ga}^{(1)} = 0.088(4)$ and $E_{Ga}^{(2)} = 0.073(4)$ (in eV).

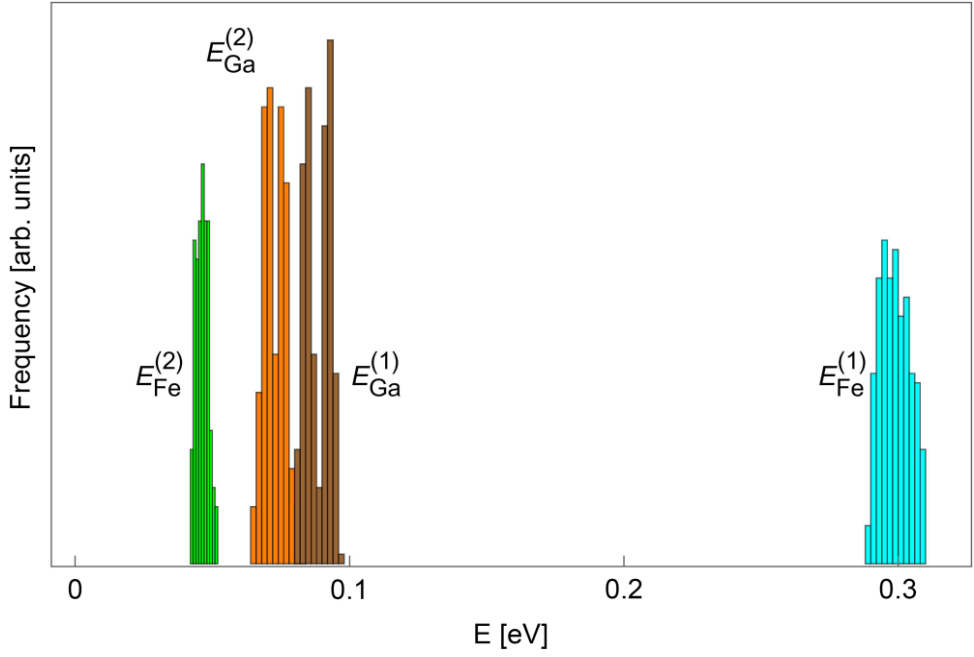

Fig. 7 Distribution of energies $E_{Fe}^{(1)}$, $E_{Fe}^{(2)}$, $E_{Ga}^{(1)}$ and $E_{Ga}^{(2)}$, see text for precise definition. One should not confuse used colors and sites Fe1, Fe2, Ga1, and Ga2.

By Mössbauer experiments (Table 8), occupation numbers were extracted (Table 9). The occupation numbers $c_{1Ga}$, $c_{2Ga}$, $c_{1Fe}$, $c_{2Fe}$ are defined as a fraction of foreign atoms occupying Fe1, Fe2, Ga1, and Ga2 sites, respectively. Note that the index in the occupation numbers shows foreign atoms located at the site indicated by the integer. All occupation numbers are between 0 and 1; the condition of stoichiometry demands that $c_{1Ga} + c_{2Ga} - c_{1Fe} - c_{2Fe} = 0$. For perfectly ordered GaFeO₃, all the occupation numbers are equal to zero. For the sake of clarity, we quote that the site Fe1 consists of $1 - c_{1Ga}$ Fe atoms and $c_{1Ga}$ foreign Ga. The occupation numbers at thermal equilibrium can be obtained using the estimated energies and statistical approach. Assuming that in $N$ positions of the given site (say, Fe1), there are $Nc_{1Fe}$ foreign Ga atoms, the number of possible microstates is equal to the number of all different $Nc_{1Fe}$-element subsets taken from the set of $N$ elements. Thus, the number of microstates $\Gamma$ for four sites with occupation numbers $c_{1Ga}, c_{2Ga}, c_{1Fe}, c_{2Fe}$ is

$$\Gamma = \binom{N}{Nc_{1Ga}}\binom{N}{Nc_{2Ga}}\binom{N}{Nc_{1Fe}}\binom{N}{Nc_{1Fe}}, \quad (7)$$

where $\binom{n}{k}$ is $n!/(k!(n-k)!)$. Using the Stirling formula for $n!$ the entropy $S = k_B \ln \Gamma$ can be obtained from (7) in the limit of large $N$:

$$S = k_B N \ln f(c_{1Ga}) f(c_{2Ga}) f(c_{1Fe}) f(c_{2Fe}), \quad (8)$$

where

$$f(x) = \frac{1}{x^x(1-x)^{(1-x)}}. \tag{9}$$

The energy $U$ is equal to

$$U = N\left(E_{Fe}^{(1)} c_{1Fe} + E_{Fe}^{(2)} c_{2Fe} + E_{Ga}^{(1)} c_{1Ga} + E_{Ga}^{(2)} c_{2Ga}\right). \tag{9}$$

The number of foreign atoms has to fulfill the chemical composition of the crystal,

$$c_{1Ga} + c_{2Ga} - c_{1Fe} - c_{2Fe} = 0. \tag{10}$$

From eq. (9) and (10) one can calculate $c_{1Fe}$ and $c_{2Fe}$. In the thermodynamic equilibrium, the entropy achieves maximum value, so the derivatives of (8) over $c_{1Ga}$ and $c_{2Ga}$ should vanish:

$$\frac{\partial S(c_{1Ga}, c_{2Ga}, U)}{\partial c_{1Ga}} = 0, \quad \frac{\partial S(c_{1Ga}, c_{2Ga}, U)}{\partial c_{2Ga}} = 0. \tag{11}$$

Eq. (11) allow us to find entropy as a function $c_{1Ga}$, $c_{1Ga}$ at given energy $U$. Further on, since $T = \partial U/\partial S$, one can find all the occupation numbers numerically $c_{1Ga}, c_{2Ga}, c_{1Fe}, c_{2Fe}$ as a function of equilibrium temperature, shown in Fig. 8a and, more conveniently, Fe occupation of four sites of GaFeO$_3$ (see Fig. 8b).

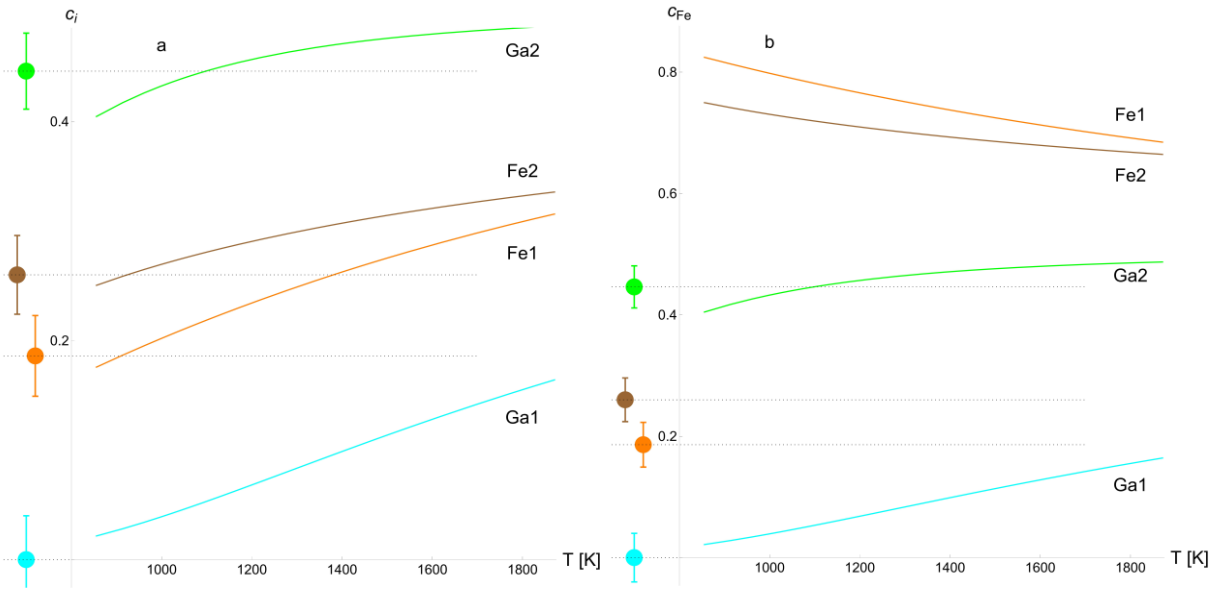

Fig. 8 a) Foreign atom site occupancies (related to occupation numbers in the third column of Table 8) and b) iron site occupancies. Both panels show, in fact, the same data. The lines show predictions of thermal equilibrium site occupancies, while points on the left of each panel show results obtained from Mössbauer experiments. Experimental values do not correspond to the thermal equilibrium state.

For the occupancies shown in Table 9 and $P_z > 0$, the crystal energy of the asymmetric unit is estimated as 0.057(13) eV, while for $P_z < 0$, 0.170(7) eV. These values correspond to the change of energy in switching electrical polarization of GaFeO$_3$ as large as $31(4)$ kJ $\cdot$ kg$^{-3}$ or 1.1(1) kJ/mol of atoms. The first value may be compared with the heat fusion of elements, gallium: 80 kJ $\cdot$ kg$^{-3}$ or lithium: 3 kJ/mol and show that the predicted switching energy is unphysically large.

Table 9. Site composition derived from the Mössbauer experiment and occupation numbers defined as fractions of foreign atoms. The site compositions (column 2) are proportional to the $ARE$ presented in Table 8, col Sub1, Sub2, Sub3: 2·(37.0, 40.7, 22.3)/100=(0.740, 0.814, 0.446).

| site | site composition Fe/Ga | occupation numbers |
|---|---|---|
| structure $P_z > 0$ | | |
| Fe1 | 0.740/0.260 | $c_{1Ga} = 0.26(1)$ |
| Fe2 | 0.814/0.186 | $c_{2Ga} = 0.19(4)$ |
| Ga1 | 0.000/1.000 | $c_{1Fe} = 0.00(4)$ |
| Ga2 | 0.446/0.554 | $c_{2Fe} = 0.45(1)$ |
| structure $P_z < 0$ | | |
| Fe1 | 0.814/0.186 | $c_1 = 0.19(4)$ |
| Fe2 | 0.740/0.260 | $c_2 = 0.26(1)$ |
| Ga1 | 0.446/0.554 | $c_3 = 0.45(1)$ |
| Ga2 | 0.000/1.000 | $c_4 = 0.00(4)$ |

The overall agreement of the combined data analysis, based on theoretical predictions and spectra analysis, guarantees correct spectra assignment. Measurements were done on different orientations of single crystal, at magic angle geometry and in-magnetic fields with a consistent set of parameters (Fig. 3 and 4). It also allows for the determination of spectral areas and quantitative estimation of the atomic disorder. Theoretical calculations of some atomic configuration conditions allow energy estimations related to the atomic disorder.

We have observed the switching of the polarization state by piezoresponse force microscopy. The measurements indicate symmetric, nonbiased hysteresis of the polarization switching. Thus the microscopic mechanism proposed in [10] can be questioned because of the difference in the crystal energy between polarization states $31(4)$ kJ $\cdot$ kg$^{-3}$ is expected.

The new concept of atomic displacements in the unit cell guarantees equal energy for two polarization states for disordered GaFeO$_3$ crystals and is consistent with observed symmetric energy switching induced by PFM spectroscopy.

## 9. Conclusions

We have synthesized GaFeO$_3$ single crystals. The site disorder was determined by X-ray diffraction and Mössbauer experiments. The microscopic energies of disorder were estimated. The location of iron in the Ge1 tetrahedral site is energetically unfavorable. This substitution has energy larger by about 0.2 meV than the other three possible types of substitution (Fe in Ga2, Ga in Fe1, and Ga in Fe2). Electronic structure calculations indicate that the hyperfine parameters of Fe in the Ga1 site are close to that of Fe1. This is a possible reason for observing only three distinct components in the Mössbauer experiments performed so far. Switching of electrical polarization was demonstrated by PFM spectroscopy. A new mechanism of electrical polarization switching of disordered GaFeO$_3$, consistent with physical properties measured so far, was proposed. In contrast to earlier concepts [10], the proposed mechanism preserves atomic disorder in the cationic sites and guarantees symmetric hysteresis of the electrical polarization switch.

**Acknowledgments**

This work was partially supported by the National Science Centre (grant OPUS no 2018/31/B/ST3/00279). K.J.K. thanks Przemysław Piekarz and Andrzej Ptok for constructive discussions and sincerely acknowledges the hospitality of the Henryk Niewodniczański Institute of Nuclear Physics (of the Polish Academy of Sciences) in Kraków (Poland) as well as access to their computational resources. K.J.K. thanks the Polish National Agency for Academic Exchange for funding in the frame of the Bekker program (PPN/BEK/2020/1/00184).